\documentclass[amssymb, amsmath, 10pt, twocolumn, nofootinbib, preprintnumbers, a4paper]{revtex4}

\begin{document}

\preprint{ULB-TH/03-09}

\title{CP violation in weak interactions from orbifold reduction:\\  possible unification structures.}

\author{N.Cosme, J.-M. Fr\`ere.}
\thanks{ncosme@ulb.ac.be, frere@ulb.ac.be.}
\affiliation{Service de Physique Th\'eorique, CP225
Universit\'e Libre de Bruxelles,
Bld du Triomphe, 1050 Brussels, Belgium.}

\begin{abstract}
We present a mechanism to generate complex phases from real $4+1$ dimensional couplings in a 
model of weak interactions through dimensional reduction of a gauge theory. The orbifolding of a $4+1$ 
dimensional $Sp(4) \times U(1)$ group is the minimal setup which provides both $CP$ violation and an $SU(2)\times U(1)$
structure. We show that grand unification requires at least $SO(11)$.
\end{abstract}

\maketitle

$CP$ violation in the standard model, since gauge interactions are naturally $CP$ symmetric, is provided by complex Yukawa 
couplings which eventually are combined in the CKM matrix in one observable $CP$ violating phase.
While this picture has been comforted through B-decay observations \cite{BabarBelle}, the standard model does not tell 
us more on the origin of $CP$ violation since it is explicitely introduced.

\medskip
On the other hand, a truly unified theory would relate Yukawa couplings to gauge interactions
implying that this unified theory would be $CP$ symmetric. In that context a $CP$ breaking mechanism is needed
and can be found, as addressed here, in dimensional reduction. One example of these possibilities has been
studied in \cite{Cosme:2002zv}. We present here a realistic realisation of these ideas in the standard model.

\medskip

In the context of five dimensional gauge theory, the reduction from $4+1$ to $3+1$ dimensions has to deal with the 
extra contribution to the energy coming from the extra component of the covariant derivative, that is: 
$D_y= \partial_y - i e A_y$, where the derivative leads to the well known Kaluza-Klein(KK) effective mass $\frac{n}{R}$ 
in $3+1$ dimensions.

For spinors, this contribution is associated to the usual $3+1$ dimensional pseudoscalar: 
$\bar{\psi} \gamma_5 \psi$, since the Clifford algebra is extended to $\gamma_B=(\gamma_\mu, i\gamma_5)$ for $4+1$ dimensions 
( $B=0,1,\cdots,(4=y)$).
Thus, whatever the reduction scheme is, the fermionic mass term may receive effective complex masses of the type:
$$\bar{\psi} (M+i \gamma_5 X) \psi.$$
This effective complex mass will lead to $CP$ violation( although in a pure minimal-coupling $U(1)$ theory the complex 
phase can be rotated away by a redefinition of spinors).

\medskip
Several contributions can be considered for $X$, e.g. the KK mass $\frac{n}{R}$  
combined with a non-minimal coupling to the photon has been studied by Thirring \cite{Thirring:1972de}.
Otherwise, in order to distinguish $CP$ violation from the use of exited states, some vacuum expectation
 value for the extra component of the gauge field, that is the gauge invariant line integral 
$$ X =\langle A_y \rangle=\int dy A_y, $$
together with an extention of the gauge group has been considered in \cite{Cosme:2002zv}. This line integral
keeps $3+1$ dimensional Lorentz invariance and reduces to the usual Wilson loop in the case of a compact extra space.

\medskip
For instance, consider a $4+1D$ $SU(2)$ gauge group with massive doublet $\Psi=\left( \psi_1, \psi_2 \right)$, and
take the expectation value $\langle W_y \rangle =\int dy \; W_y =w \; \sigma^3$ to break the group to vectorlike 
effective interactions in $3+1$ dimensions:
$$\left(\begin{array}{cc}\bar{\psi_1} & \bar{\psi_2}\end{array}\right)i(\partial^\mu -i W^\mu_a \tau^a) \gamma_\mu
 \left(\begin{array}{c}\psi_1 \\ \psi_2\end{array}\right),$$
with two massive $W^\pm$ and one massless $W^3$. Then, the Wilson loop contributes to a complex mass matrix:
$$\left(\begin{array}{cc}\bar{\psi_1} & \bar{\psi_2}\end{array}\right)
\left(\begin{array}{cc} M+iw\gamma_5&  \\  &M-iw\gamma_5\end{array}\right)
 \left(\begin{array}{c}\psi_1 \\ \psi_2\end{array}\right).$$

Both phases cannot be redefined and, while making the mass matrix real, a remaining phase appears in the charge 
current implying a $W^3$-dipole moment at one loop level, i.e. a $CP$ violating observable.

This example shows that in this approach, realisation of $CP$ violation, dimensional reduction and breaking of the internal symmetry are intimately related.
Moreover, $CP$ violation is generated in a fundamentally $CP$ symmetrical framework where all initial couplings are real. 

\medskip
In this, the approach differs from \cite{Branco:2000rb} where the $CP$ violation is explicitely introduced and 
\cite{Chang:2001yn} where the $LR$ violation stems from dimensional reduction, but scalar couplings are localized
in $3+1$ dimensions. The line followed here is similar to \cite{Cosme:2002zv} where we had dealt only with toy
gauge structure and a simple compactified extra dimension. The use of the 5th component of a gauge vector to 
provide scalar couplings is very much in the line of Kaluza-Klein tradition (see also \cite{Gogoladze:2003bb}),
but it is used here to generate specifically the $CP$ violation in an otherwise real theory.

\medskip
In $4+1$ dimensions, only "vectorlike" couplings arise since chirality does not exist. So, as weak
interactions are intrinsically chiral, the reduction scheme has to introduce a selection of chirality. Nevertheless,
since our goal is to form mass terms through the gauge Wilson loop, the breaking of the symmetry should
keep some L and R components. We thus choose a reduction scheme which selects as many left- as right-handed fermions.

It results that the initial theory should contain the minimal left-right extention of weak interactions, that is
$SU(2)_L \times SU(2)_R \times U(1)_{B-L}$. Gauge groups containing this left-right structure are, e.g. $SU(4)$,
$Sp(4)$, etc. 


\section{Orbifold Reduction.}

Orbifolds provide a breaking of higher dimensional symmetries ( such as chiral, super or gauge symmetries) via an internal
geometric symmetry of the extra space. This geometric symmetry induces a transformation on the fields and selects zero modes
 which break the higher dimensional invariance \cite{Hebecker:2001jb}. 

More explicitely, in $4+1$ dimensions, we take the extra space dimension as a $S^1/Z_2$, i.e. the circle with 
the points identification under the $4+1D$ parity( $y \rightarrow -y$). This fixes the geometric space we work with.

Moreover, we have to specify the $Z_2$ representation on the field content. Actually, for any transformation under parity 
of the Lorentz representations, we are allowed to add in the transformation a symmetry of the theory, for instance
a gauge transformation $P_{\mathcal{G}} \in \mathcal{G}$ (with $P_{\mathcal{G}}^2 =\Bbb{I}$). 

So we get:
\begin{eqnarray}
 A_\mu^a (x_\nu,y)\; \lambda^a & = & A_\mu^a (x_\nu,-y) \; P_{\mathcal{G}} \lambda^a P_{\mathcal{G}}^{-1},\nonumber\\
 A_y^a (x_\nu,y) \;\lambda^a  &= & - A_y^a (x_\nu,-y) \; P_{\mathcal{G}} \lambda^a P_{\mathcal{G}}^{-1}, \nonumber
\end{eqnarray}
for gauge fields and
$$\Psi(x_\mu,y)  = P_{\mathcal{G}} \gamma_5   \Psi(x_\mu,-y),$$
for fermions.

This identification determines the KK expansion for fields with respect to their parity eigenvalues:
$\cos{ \frac{n}{R}y}$ for $+1$ while $\sin{\frac{n}{R}y}$ for $-1$. 

Subsequently, $P_{\mathcal{G}}$ can be chosen to commute with the $\lambda^{\bar{a}}$ generating the $\mathcal{\bar{G}}$ subgroup 
while anticommuting with the other $\lambda^{\hat{a}}$ ($a=\left( \bar{a}, \hat{a} \right)$). In that
way, zero modes $A_\mu^{\bar{a},  0}$ belong to an unbroken $\mathcal{\bar{G}} \subset
\mathcal{G}$. On the other hand, the zero modes for the extra component of the gauge fields
are the $A_y^{\hat{a}, 0}$ and belong to the coset $\mathcal{G}/\mathcal{\bar{G}}$. 

Fermion zero modes depend on both the sign of chirality and the sign from the gauge transformation $P_{\mathcal{G}}$. This
implies that the initially vectorlike fermionic representation is then split in chiral representations
under the unbroken group $\mathcal{\bar{G}}$. L and R representations are coupled through the Wilson loop
to form a complex mass.

For scalars now, there are two cases.\footnote{We just focus here on adjoint scalars.} First, if they couple
to fermions, since the $\bar{\psi}\psi$ term is not invariant under $Z_2$ \cite{Gavela}, the identification must be:
$$ \Phi^a (x_\mu,y) \; \lambda^a  =  - \Phi^a (x_\mu,-y) \;  P_{\mathcal{G}} \lambda^a P_{\mathcal{G}}^{-1}. $$
Zero modes are then in $\mathcal{G} / \mathcal{\bar{G}}$ and their vev's are aligned to $\int dy A_y$
to minimize the interaction potential coming from the covariant derivative.\footnote{Note that, if we want a $CP$ 
invariant mass term, we have also to add a sign for the charge conjugate of such a scalar: 
$ \mathcal{C} \Phi \mathcal{C} = - \Phi^c$.} Otherwise for scalars not directly coupled to fermions,
the sign of the transformation is free.

The obvious advantage of any orbifold rather than physical domain wall is the purely geometrical approach.
This does not lead to any problem of stabilisation for the domain wall nor localisation of gauge fields.
It also avoids parasitic solutions as in domain wall approach on compactified spaces.


\section{Minimal model.}
Now, we start with a gauge theory in $4+1$ dimensions which will reduce to a $3+1$ left-right symmetric
gauge theory with complex Yukawa couplings. Let first recall the basic fields for an $SU(2)_L \times SU(2)_R \times U(1)_{B-L}$
model:

\begin{center}
\begin{tabular}{|c||c|}
\hline  & $SU(2)_L \times SU(2)_R \times U(1)_{B-L}$\\
\hline \hline $Q_L $&  $(2,1)_{(B-L)}$\\
\hline  $Q_R$&  $(1,2)_{(B-L)}$\\
\hline  $\Phi$& $ (2,2)_{0}$\\
\hline $\chi_L $& $ (3,1)_{2} $ or $(2,1)_{1}$\\
\hline $\chi_R$&  $(1,3)_{2} $ or $(1,2)_{1}$\\
\hline
\end{tabular}
\end{center}

The bi-doublet $\Phi$ breaks both $SU(2)$'s leaving $ U(1)_{T^3_R+T^3_L} \times U(1)_{B-L}$ unbroken. $\chi_{L,R}$
differentiate both $SU(2)$'s with their respective vev's giving a mass $\mathcal{O}(\langle \chi_R \rangle)$ to the $W_R$.
We list also the maximal subgroups of $SU(4)$ and $Sp(4)$, and the representation decomposition below:
\footnote{The notation is self-explanatory: dimension of the representation for $SU$ components and $U(1)$ charge
for the abelian part.}

\begin{center}
\begin{tabular}{c|l}
$ Sp(4) $& $\supset SU(2) \times SU(2) $ \\
\hline $4$\text{ (fundamental)}& $\rightarrow (2,1)+(1,2) $\\
 $10$\text{  (adjoint)}&$ \rightarrow (3,1)+(1,3)+(2,2) $\\

\end{tabular}
\end{center}
\begin{center}
\begin{tabular}{c|l}
$ SU(4)$ & $\supset SU(2) \times SU(2) \times U(1)$ \\
\hline
 $4$\text{  (fundamental)} & $\rightarrow (2,1)(1)+(1,2)(-1) $\\
 $15$\text{  (adjoint)}& $ \rightarrow (1,1)(0)+(3,1)(0)+(1,3)(0)$ \\
   & $ +(2,2)(2)+(2,2)(-2)$  \\
\end{tabular}
\end{center}


First consider an $SU(4)$ gauge group and the parity operator $P_{\mathcal{G}}=diag(1,1,-1,-1)$ acting on the fundamental.
We verify easily that $P_{\mathcal{G}}$ commutes with  generators of an $SU(2) \times SU(2) \times U(1)$ subgroup while
anticommuting with the others.

As a result, the gauge zero modes are:
\begin{eqnarray}
 A_\mu^{\bar{a},0} & \rightarrow &(1,1)(0)+(3,1)(0)+(1,3)(0), \nonumber \\
A_y^{\hat{a},0} & \rightarrow & (2,2)(2)+(2,2)(-2),\nonumber
\end{eqnarray}
and fermions in the $4$ representation reduce to the following zero modes:
$$\left(\begin{array}{cccc} u^0_L & d^0_L  & u^0_R  & d^0_R \end{array}\right) \leftrightarrow (2,1)(1)+(1,2)(-1).$$
An adjoint scalar $\Phi$ coupled to fermions gets its zero modes in the same representation as $A_y$.

Could we get the same building blocks as in the left-right model? On one hand, we indeed get left and right doublets
for fermions together with a bi-doublet for both $\Phi$ and $A_y$. But on the other hand, since the obtained $U(1)_X$ 
differentiates left and right fermions, we are not able to use it as a $U(1)_{B-L}$. The alternative is then to start
from $SU(4) \times U(1)$ and to assign ourselves the hypercharges to the representations. We should eventually
care to break the unwanted $U(1)_X$ at least at the same level than $SU(2)_R$ in order to eliminate it from the
low energy spectrum. To do that, $\chi_{L,R}$ cannot be put in the adjoint since it reduces to $(3,1)(0)+(1,3)(0)$ without
giving a mass to $U(1)_X$. The remaining possibility is to put them in the $4$.

\bigskip
Therefore, the field content of the theory is:

\noindent
\begin{tabular}{|c|c||c|}
\hline   & $ SU(4) \times U(1)_{B-L} $ & $SU(2)_L \times SU(2)_R$\\
  & & $\times U(1)_X \times U(1)_{B-L}$ \\ 
\hline
\hline  $B_\mu$ & $1_0$ & $  (1,1)(0)_0$  \\ 
\hline $A_\mu$ & $15_0$ & $(3,1)(0)_0+(1,3)(0)_0+(1,1)(0)_0$\\ 
      $A_y $& $15_0$ &  $(2,2)(2)_0+(2,2)(-2)_0$\\ 
\hline $ \Psi $& $4_{(B-L)} $&  $(2,1)(1)_{(B-L)}+(1,2)(-1)_{(B-L)}$\\ 
\hline $ \Phi$ & $15_0 $&  $(2,2)(2)_0+(2,2)(-2)_0$\\ 
\hline $ \chi_{L,R}$ &  $ 4_1$ &  $(2,1)(1)_{1}+(1,2)(-1)_{1}$ \\
\hline 
\end{tabular}
\medskip

This reproduces a left-right model with one more neutral current with mass $\mathcal{O}(\langle \chi_R \rangle)$.
Complex phases are obtained in the fermion mass matrix:
$$ \bar{u}^0_L \;\;\frac{1}{2}(m \langle \Phi_4 \rangle +i \langle A^y_4 \rangle  \gamma_5) \;\;u^0_R $$
$$+ \bar{d}^0_L \;\;\frac{1}{2}(m \langle \Phi_{11} \rangle +i  \langle A^y_{11} \rangle \gamma_5)\;\; d^0_R + h.c.. $$

CP violation occurs with both $W_L$ and $W_R$ interactions and only one generation. However, to get $CP$ violation 
through the $W_L$ alone, more generations are needed to form the usual CKM matrix. 

Note that $CP$ violation is induced 
here at the dimensional reduction stage, not at the level of $LR$ breaking which thus avoids difficulties met in 
\cite{Ball:1999mb}.

\medskip

The $Sp(4)$ case seems more attractive since it contains $SU(2) \times SU(2)$ without any other $U(1)$. So, starting
from $Sp(4) \times U(1)_{B-L}$ with parity $P_{\mathcal{G}}=diag(1,-1,1,-1)$,\footnote{our conventions for $Sp(4)$ are listed in 
appendix \ref{SP4}} the group breaks down to $SU(2)_L \times SU(2)_R \times U(1)_{B-L}$. The field content:
\begin{center}
\begin{tabular}{|c|c||c|}

\hline   & $ Sp(4) \times U(1)_{B-L}$ & $SU(2)_L \times SU(2)_R \times U(1)_{B-L}$  \\ 
\hline
\hline  $B_\mu$ & $1_0$ & $  (1,1)(0)_0$  \\ 
\hline $A_\mu$ & $10_0$ & $(3,1)_0+(1,3)_0$\\ 
       $A_y $& $10_0$ &  $(2,2)_0 $\\ 
\hline $ \Psi $& $4_{(B-L)} $&  $(2,1)_{(B-L)}+(1,2)_{(B-L)}$\\ 
\hline $ \Phi$ & $10_0 $&  $(2,2)_0 $\\ 
\hline $ \chi_{L,R}$ & $ 10_2$ &  $(3,1)_2+(1,3)_2$ \\ 
\hline 
\end{tabular}
\end{center}
provides the desired breaking pattern and realises the minimal
requirements for a realistic model. $CP$ violation arises as in the previous case.


\section{Unification.}
We now discuss possible embeddings of such a model in a unique gauge group which contains both strong and
electroweak interactions. 


Since the structure is left-right symmetric, the first group which could potentially be considered
is $SO(10)$. Indeed, $SO(10)$ contains maximally $SU(2)_L \times SU(2)_R \times SU(4)$, where $SU(4)$ can be broken 
to $SU(3)_c \times U(1)_{B-L}$\cite{Ross:ai}. 

\smallskip
Nevertheless, the fermions unification in $SO(10)$ comes with the 16 representation which in $3+1D$
includes only left fermions, e.g. $( \left( u_L, d_L \right) \; \cdots \; \left(u^c_L,d^c_L\right) )$. Therefore,
as chirality cannot be assigned to representations in $4+1D$ and as we need L and R fermions to come out from
the same representation to get a Wilson loop coupling, the usual unification is not appropriate here.

At least, if we replace the left handed anti-particles by their corresponding right handed particles, 
charges under the resulting $U(1)$ and $SU(3)$ will differentiate L and R fermions.
 Indeed, the reduction of the 16 is:
$16 \rightarrow (2,1,4) +(1,2, \bar{4})$, where the $4$ reduces in $4 \rightarrow 1_3 +3_{-1}$. This is however incompatible
with the fermion spectum as we then get quarks as $Q_L^i: (2,1,3)_{-1}$ and $Q_R^i: (1,2,\bar{3})_{1}$. The $U(1)$ charge 
is as in the $SU(4)$ case but moreover the quark mass term $Q^\dagger_L Q_R$ is no longer an $SU(3)$ singlet and then breaks $SU(3)$.

In other words, the $16$ can be reduced either to 
$( \left( u_L, d_L \right) \;\cdots \; \left(u^c_L,d^c_L\right) ),$
with $P_{\mathcal{G}}= \Bbb{I}$, or
$( \left( u_L, d_L \right) \;\cdots \; \left(u_R,d_R\right) ).$
Only the first choice gives the right particle content but it is then impossible to generate 
the desired Yukawa couplings in the present scheme.
\medskip

We will now see that $SO(11)$ answers those problems. Indeed, the doubled fermion components resolve the chirality problem.

Since $SO(11)$ is not such a common unification group, we first consentrate on the group structure and 
decomposition before dealing with the reduction itself.

$SO(11)$ obviously contains $SO(5)\times SO(6)$, that is up to an isomorphism $Sp(4)\times SU(4)$.
As already said, $Sp(4)$ can provide the left-right extension for weak interactions while $SU(4)$ is often used in the
more usual $SO(10)$ to get strong interactions. From the point of view of representations, the $SO(11) \to Sp(4)
\times SU(4)$ breaking induces the following reduction \cite{Slansky:yr}:
\begin{eqnarray}
32 \mbox{ (spin$\frac{1}{2}$)} &\to& (4,4)+(4, \bar{4}), \nonumber \\
55 \mbox{ (adjoint)} &\to &(10,1)+(1,15)+(5,6).\nonumber 
\end{eqnarray}

As a next step, $Sp(4) \to SU(2)_L \times SU(2)_R$ and $SU(4) \to SU(3)_c \times U(1)_{B-L}$ breakings imply 
the spectrum:
\begin{eqnarray}
(4,4) &\to &(2,1,1)_3 +(1,2,1)_3 +(2,1,3)_{-1}+(1,2,3)_{-1}, \nonumber \\
(10,1) &\to & (3,1,1)_0+(1,3,1)_0+(2,2,1)_0, \nonumber \\
(1,15) &\to & (1,1,1)_0+(1,1,8)_0+(1,1,3)_{-4}+(1,1,\bar{3})_{4}.\nonumber 
\end{eqnarray}

Thus, if we ensure that the first breaking does not select chirality while the second does, fermions in the $32$ reduce to a 
vectorlike $(4,4)$ of $Sp(4) \times SU(4)$ which gives rise to an entire fermion family with the correct charges and 
chiralities.\footnote{the $(4,\bar{4})$ being eliminated by orbifolding, see below.} 
\smallskip

Let us now turn to the dimensional reduction of this $SO(11)$ compactified on a $S^1/Z_2$.

Since the first breaking has to be left-right blind, it won't result from a $Z_2$ symmetry along the extra coordinate. 
However, another possibility is offered to us, that is to allow a gauge transformation in the $S^1$ periodic conditions.
Indeed, extending periodic conditions to:
$$\Psi(y+2\pi R) = T_{\mathcal{G}} \; \Psi(y),$$
$$A^a_B(y+2\pi R) \lambda^a = A^a_B(y) \;  T_{\mathcal{G}} \lambda^a T^{-1}_{\mathcal{G}},$$
selects zero modes with eigenvalues $+1$. Since for $+1$ the KK tower contains both $\cos{\frac{n}{R}y}$ and 
$\sin{\frac{n}{R}y}$ while for $-1$ the complete set of functions are $\cos{(n+\frac{1}{2})\frac{y}{R}}$ and 
$\sin{(n+\frac{1}{2})\frac{y}{R}}$, the latter's have no zero modes.

So, in that way, we take 
$$T_{\mathcal{G}}=\left(\begin{array}{cc} \Bbb{I}_{5\times 5} & 0 \\ 0 &-\Bbb{I}_{6\times 6} \end{array}\right)$$
in the fundamental of $SO(11)$, i.e. the inversion of the fundamental of $SO(6)$. This selects zero modes for the adjoint of $SO(11)$ in $(10,1)+(1,15)$ of the 
unbroken group. The $32$ gets its zero modes in the $(4,4)$.\footnote{see appendix \ref{SO6} for clarity.}

The second breaking takes place with the $Z_2$ symmetry where $P_{\mathcal{G}}$ has to be determined. 
The $Sp(4)$ part has already been considered before. 
The $SU(4)$ part however cannot be broken to $SU(3)\times U(1)$ through a $Z_2$ orbifold since
there is no automorphism to play that role \cite{Hebecker:2001jb}.\footnote{Roughly, this is due to the requirement of
a $det=+1$ transformation.}
Nevertheless, this breaking can be provided by an adjoint scalar of $SO(11)$, not coupled to fermions, which gets a vev in the $(1,1,1)_0$ representation
of $SU(2)_L\times SU(2)_R\times SU(3)_c \times U(1)_{B-L}$. This indeed breaks $SU(4)$ to $SU(3)\times U(1)$.

Thus, the gauge transformation parity $P_{\mathcal{G}}$ takes the form of $diag(1,-1,1,-1)$ for the $Sp(4)$ part in direct product with
the identity for $SU(4)$. 

In the same way as before, the $Sp(4)$ group gives rise to the left-right $SU(2)_L \times SU(2)_R$,
with this symmetric structure for fermions and bi-doublets for a mass scalar and the Wilson loop.
The last ingredients for this symmetry to be broken are two $\chi_{L,R}$ in the $32$ which transform
as:
$$\chi_{L,R}(y+2\pi R)=-T_{\mathcal{G}} \; \chi_{L,R}(y),$$
under periodic conditions and which respectively get their vev in the $(2,1,1)_{-3}$ and $(1,2,1)_{-3}$. 

We summarize below the cascade with the needed breaking sector:
$$SO(11) \stackrel{T_{\mathcal{G}}}{\rightarrow} Sp(4) \times SU(4) \stackrel{P_{\mathcal{G}}}{\rightarrow} SU(2)_L \times SU(2)_R \times SU(4) 
\stackrel{55}{\rightarrow} $$
$$SU(2)_L \times SU(2)_R \times SU(3)_c \times U(1)_{B-L} 
\stackrel{55 \, \& \, 32}{\rightarrow} SU(3)_c \times U(1)_{Q}$$


\section{Conclusion}
We have explored group structures in $4+1$ dimensions which either reproduce the standard model
(using $Sp(4)$) or allow for grand unification via the left-right symmetric model.

At the difference of standard left-right model, $CP$ violation is here present already at the compactification 
scale, before left-right breaking.

Of courses the three generations still need to be introduced by hand, as the real Yukawa couplings needed 
to define the mass spectrum. We have achieved here a mechanism for breaking $SO(11)$ to the standard model, 
and generate the $CP$ violating part of the couplings.

\appendix

\section{$Sp(4)$ generators.}\label{SP4}
We list for completeness $Sp(4)$ generators used here.
$$
T_i^L = \sigma_i \times S^L, \;\;
T_i^R = \sigma_i \times S^R, \;\;
T_i^3 = \sigma_i \times S^3, \;\;
T_{10}  = \Bbb{I} \times A ; 
$$
where:
$$ 
S^L=\left( \begin{array}{cc}
1 & 0 \\ 
0 & 0
\end{array}\right) , \qquad 
S^R = \left( \begin{array}{cc}
0 & 0 \\ 
0 & 1
\end{array}\right) , 
$$
$$
S^3=\frac{1}{\sqrt{2}} \left( \begin{array}{cc}
1 & 0 \\ 
0 & 1
\end{array}\right)  , \qquad
A= \frac{1}{2\sqrt{2}}\left( \begin{array}{cc}
0 & -i \\ 
i & 0
\end{array}\right),  
$$
and $\sigma_i$ are the Pauli matrices.

\section{$4$ and $\bar{4}$ of $SO(6)$}\label{SO6}

The spinorial representations of $SO(6)\; (\equiv SU(4))$ is given by the Clifford algebra of six $\Gamma$ matrices:
$$\Gamma_1 =\sigma_2 \times \sigma_3 \times \sigma_3, \qquad \Gamma_2 =-\sigma_1 \times \sigma_3 \times \sigma_3,
\qquad \Gamma_3 = \Bbb{I} \times \sigma_2 \times \sigma_3, $$
$$\Gamma_4 = -\Bbb{I} \times \sigma_1 \times \sigma_3, \qquad \Gamma_5 = \Bbb{I} \times \Bbb{I} \times \sigma_2,
\qquad \Gamma_6 = -\Bbb{I} \times \Bbb{I} \times \sigma_1,$$
which provide the generators of the representation, i.e. $M_{ij}=\frac{1}{4i} [\Gamma_i,\Gamma_j]$.
Moreover, this representation is reducible in a $4$ and a $\bar{4}$, with the projectors given by:
$\frac{1}{2} (\Bbb{I} \pm \Gamma_7)$, where $\Gamma_7= \sigma_3 \times \sigma_3 \times \sigma_3$ \cite{Georgi:jb}.

Since the set of $\Gamma_i$'s transforms as the fondamental of $SO(6)$, it is easy to check that $\Gamma_7$ is the equivalent
of the inversion of the fundamental of $SO(6)$ and to observe that the $4$ is unchanged while the $\bar{4}$ takes
a minus sign.
\section*{Acknowledgments.}

We thank  M. Tytgat and Y. Gouverneur for interesting discussions. This work is 
supported in part by IISN, la Communaut\'{e} Fran\c{c}aise de Belgique (ARC), and the belgian 
federal government (IUAP).



\begin{thebibliography}{99}
\bibitem{BabarBelle}
B. Aubert \textit{et al.} [Babar collaboration], Phys. Rev. Lett. {\bf 87},
091801 (2001) [hep-ex/0107013]. K. Abe \textit{et al.} [Belle Collaboration], 
Phys. Rev. Lett. {\bf 87}, 091802 (2001) [hep-ex/0107061].

\bibitem{Cosme:2002zv}
N.~Cosme, J.~M.~Frere and L.~Lopez Honorez,
Phys.\ Rev.\ D {\bf 68} (2003) 096001
[arXiv:hep-ph/0207024].


\bibitem{Thirring:1972de}
W.~E.~Thirring,
Acta Phys.\ Austriaca Suppl.\  {\bf 9} (1972) 256.

\bibitem{Branco:2000rb}
G.~C.~Branco, A.~de Gouvea and M.~N.~Rebelo,
Phys.\ Lett.\ B {\bf 506} (2001) 115
[arXiv:hep-ph/0012289].

\bibitem{Chang:2001yn}
D.~Chang and R.~N.~Mohapatra,
Phys.\ Rev.\ Lett.\  {\bf 87} (2001) 211601
[arXiv:hep-ph/0103342].
D.~Chang, W.~Y.~Keung and R.~N.~Mohapatra,
Phys.\ Lett.\ B {\bf 515} (2001) 431
[arXiv:hep-ph/0105177].

\bibitem{Gogoladze:2003bb}
I.~Antoniadis and K.~Benakli,
Phys.\ Lett.\ B {\bf 326} (1994) 69
[arXiv:hep-th/9310151].
C.~Csaki, C.~Grojean and H.~Murayama,
arXiv:hep-ph/0210133.
G.~Burdman and Y.~Nomura,
arXiv:hep-ph/0210257.
N.~Haba and Y.~Shimizu,
arXiv:hep-ph/0212166.
I.~Gogoladze, Y.~Mimura and S.~Nandi,
arXiv:hep-ph/0301014.


\bibitem{Hebecker:2001jb}
Y.~Kawamura,
Prog.\ Theor.\ Phys.\  {\bf 103} (2000) 613
[arXiv:hep-ph/9902423].
R.~Barbieri, L.~J.~Hall and Y.~Nomura,
Phys.\ Rev.\ D {\bf 63} (2001) 105007
[arXiv:hep-ph/0011311].
A.~Hebecker and J.~March-Russell,
Nucl.\ Phys.\ B {\bf 625} (2002) 128
[arXiv:hep-ph/0107039].
Q.~Shafi and Z.~Tavartkiladze,
Phys.\ Rev.\ D {\bf 66} (2002) 115002.

\bibitem{Gavela}
M.~Belen Gavela and R.~I.~Nepomechie,
Class.\ Quant.\ Grav.\  {\bf 1} (1984) L21.

\bibitem{Ball:1999mb}
P.~Ball, J.~M.~Frere and J.~Matias,
Nucl.\ Phys.\ B {\bf 572} (2000) 3
[arXiv:hep-ph/9910211] and ref. therein.

\bibitem{Ross:ai}
G.~G.~Ross,
{\it  Reading, Usa: Benjamin/cummings (1984) 497 P. (Frontiers In Physics, 60)}.

\bibitem{Slansky:yr}
R.~Slansky,
Phys.\ Rept.\  {\bf 79} (1981) 1.

\bibitem{Georgi:jb}
H.~Georgi,
Front.\ Phys.\  {\bf 54} (1982) 1.



\end{thebibliography}
\end{document}